# Pricing Credit Default Swap Subject to Counterparty Risk and Collateralization

Alan White[1]

FinPricing

**ABSTRACT**

This article presents a new model for valuing a credit default swap (CDS) contract that is affected by multiple credit risks of the buyer, seller and reference entity. We show that default dependency has a significant impact on asset pricing. In fact, correlated default risk is one of the most pervasive threats in financial markets. We also show that a fully collateralized CDS is not equivalent to a risk-free one. In other words, full collateralization cannot eliminate counterparty risk completely in the CDS market.

**Key Words**: valuation model; credit risk modeling; collateralization; correlation, CDS.

[1] Email: alan.white@finpricing.com Url: http://www.finpricing.com

## 1 Introduction

There are two primary types of models that attempt to describe default processes in the literature: structural models and reduced-form (or intensity) models. Many practitioners in the credit trading arena have tended to gravitate toward the reduced-from models given their mathematical tractability.

Central to the reduced-form models is the assumption that multiple defaults are independent conditional on the state of the economy. In reality, however, the default of one party might affect the default probabilities of other parties. Collin-Dufresne et al. (2003) and Zhang and Jorion (2007) find that a major credit event at one firm is associated with significant increases in the credit spreads of other firms. Giesecke (2004), Das et al. (2006), and Lando and Nielsen (2010) find that a defaulting firm can weaken the firms in its network of business links. These findings have important implications for the management of credit risk portfolios, where default relationships need to be explicitly modeled.

The main drawback of the conditionally independent assumption or the reduced-form models is that the range of default correlations that can be achieved is typically too low when compared with empirical default correlations (see Das et al. (2007)). The responses to correct this weakness can be generally classified into two categories: endogenous default relationship approaches and exogenous default relationship approaches.

The endogenous approaches include the contagion (or infectious) models and frailty models. The frailty models (see Duffie et al. (2009), Koopman et al. (2011), etc) describe default clustering based on some unobservable explanatory variables. In variations of contagion or infectious type models (see Davis and Lo (2001), Jarrow and Yu (2001), etc.), the assumption of conditional independence is relaxed and default intensities are made to depend on default events of other entities. Contagion and frailty models fill an important gap but at the cost of analytic tractability. They can be especially difficult to implement for large portfolios.

The exogenous approaches (see Li (2000), Laurent and Gregory (2005), Hull and White (2004), Brigo et al. (2011), etc) attempt to link marginal default probability distributions to the joint default probability distribution through some external functions. Due to their simplicity in use, practitioners lean toward the exogenous ones.

Given a default model, one can value a risky derivative contract and compute credit value adjustment (CVA) that is a relatively new area of financial derivative modeling and trading. CVA is the expected loss arising from the default of a counterparty (see Brigo and Capponi (2008), Lipton and Sepp (2009), Pykhtin and Zhu (2006), Gregory (2009), Bielecki et al (2013) and Crepey (2015), etc.)

Collateralization as one of the primary credit risk mitigation techniques becomes increasingly important and widespread in derivatives transactions. According the ISDA (2013), 73.7% of all OTC derivatives trades (cleared ad non-cleared) are subject to collateral agreements. For large firms, the figure is 80.7%. On an asset class basis, 83.0% of all CDS transactions and 79.2% of all fixed income transactions are collateralized. For large firms, the figures are 96.3% and 89.4%, respectively. Previous studies on collateralization include Johannes and Sundaresan (2007), Fuijii and Takahahsi (2012), Piterbarg (2010), Bielecki, et al (2013) and Hull and White (2014), etc.

This paper presents a new framework for valuing defaultable financial contracts with or without collateral arrangements. The framework characterizes default dependencies exogenously, and models collateral processes directly based on the fundamental principals of collateral agreements. For brevity we focus on CDS contracts, but many of the points we make are equally applicable to other derivatives. CDS has trilateral credit risk, where three parties – buyer, seller and reference entity – are defaultable.

In general, a CDS contract is used to transfer the credit risk of a reference entity from one party to another. The risk circularity that transfers one type of risk (reference credit risk) into another (counterparty credit risk) within the CDS market is a concern for financial stability. Some people claim that the CDS market has increased financial contagion or even propose an outright ban on these instruments.

The standard CDS pricing model in the market assumes that there is no counterparty risk. Although this oversimplified model may be accepted in normal market conditions, its reliability in times of distress has recently been questioned. In fact, counterparty risk has become one of the most dangerous threats to the CDS market. For some time now it has been realized that, in order to value a CDS properly, counterparty effects have to be taken into account (see ECB (2009)).

We bring the concept of *comvariance* into the area of credit risk modeling to capture the statistical relationship among three or more random variables. Comvariance was first introduced to economics by Deardorff (1982), who used this measurement to correlate three factors in international trading. Furthermore, we define a new statistics, *comrelation*, as a scaled version of comvariance. Accounting for default correlations and comrelations becomes important in determining CDS premia, especially during the credit crisis. Our analysis shows that the effect of default dependencies on a CDS premium from large to small accordingly is i) the default correlation between the protection seller and the reference entity, ii) the default comrelation, iii) the default correlation between the protection buyer and the reference entity, and iv) the default correlation between the protection buyer and the protection seller. In particular, we find that the default comvariance/comrelation has substantial effects on the asset pricing and risk management, which have never been documented.

There is a significant increase in the use of collateral for CDS after the recent financial crises. Many people believe that, if a CDS is fully collateralized, there is no risk of failure to pay. Collateral posting regimes are originally designed and utilized for bilateral risk products, e.g., interest rate swap (IRS), but there are many reasons to be concerned about the success of collateral posting in offsetting the risk of CDS contracts. First, the value of CDS contracts tends to move very suddenly with big jumps, whereas the price movements of *IRS* contracts are far smoother and less volatile than *CDS* prices. Second, CDS spreads can widen very rapidly. Third, CDS contracts have many more risk factors than *IRS* contracts. In fact, our model shows that full collateralization *cannot* eliminate counterparty risk completely for a CDS contract.

The rest of this paper is organized as follows: Pricing multilateral defaultable financial contract is elaborated on in Section 2; numerical results are provided in Section 3; the conclusions are presented in Section 4. All proofs and some detailed derivations are contained in the appendices.

## 2 Pricing Financial Contracts Subject to Multiple Credit Risks

We consider a filtered probability space ($\Omega$, $\mathcal{F}$, $\{\mathcal{F}_t\}_{t\geq 0}$, $\mathcal{P}$) satisfying the usual conditions, where $\Omega$ denotes a sample space, $\mathcal{F}$ denotes a $\sigma$-algebra, $\mathcal{P}$ denotes a probability measure, and $\{\mathcal{F}_t\}_{t\geq 0}$ denotes a filtration.

In the reduced-form approach, the stopping (or default) time $\tau_i$ of firm *i* is modeled as a Cox arrival process (also known as a doubly stochastic Poisson process) whose first jump occurs at default and is defined by,

$$\tau_i = \inf\left\{ t : \int_0^t h_i(s, Z_s)ds \geq H_i \right\} \tag{1}$$

where $h_i(t)$ or $h_i(t, Z_t)$ denotes the stochastic hazard rate or arrival intensity dependent on an exogenous common state $Z_t$, and $H_i$ is a unit exponential random variable independent of $Z_t$.

It is well-known that the survival probability from time *t* to *s* in this framework is defined by

$$p_i(t,s) := P_i(\tau > s \mid \tau > t, Z_t) = \exp\left( -\int_t^s h_i(u)du \right) \tag{2a}$$

The default probability for the period (*t*, *s*) in this framework is given by

$$q_i(t,s) := P_i(\tau \leq s \mid \tau > t, Z_t) = 1 - p_i(t,s) = 1 - \exp\left( -\int_t^s h_i(u)du \right) \tag{2b}$$

There is ample evidence that corporate defaults are correlated. The default of a firm's counterparty might affect its own default probability. Thus, default correlation/dependence occurs due to the counterparty relations.

The interest in the financial industry for the modeling and pricing of multilateral defaultable instruments arises mainly in two respects: in the management of credit risk at a portfolio level and in the valuation of credit derivatives. Central to the valuation and risk management of credit derivatives and risky portfolios is the problem of default relationship.

Let us discuss a three-party case first. A CDS is a good example of a trilateral defaultable instrument where the three parties are counterparties *A*, *B* and reference entity *C*. In a standard CDS contract one party purchases credit protection from another party, to cover the loss of the face value of a reference entity following a credit event. The protection buyer makes periodic payments to the seller until the maturity date or until a credit event occurs. A credit event usually requires a final accrual payment by the buyer and a loss protection payment by the protection seller. The protection payment is equal to the difference between par and the price of the cheapest to deliver (CTD) asset of the reference entity on the face value of the protection.

A CDS is normally used to transfer the credit risk of a reference entity between two counterparties. The contract reduces the credit risk of the reference entity but gives rise to another form of risk: *counterparty risk*. Since the dealers are highly concentrated within a small group, any of them may be too big to fail. The interconnected nature, with dealers being tied to each other through chains of OTC derivatives, results in increased contagion risk. Due to its concentration and interconnectedness, the CDS market seems to pose a systemic risk to financial market stability. In fact, the CDS is blamed for playing a pivotal role in the collapse of Lehman Brothers and the disintegration of AIG.

For years, a widespread practice in the market has been to mark CDS to market without taking the counterparty risk into account. The realization that even the most prestigious investment banks could go bankrupt has shattered the foundation of the practice. It is wiser to face frankly the real complexities of pricing a CDS than to indulge in simplifications that have proved treacherous. For some time now it has been realized that, in order to value a CDS properly, counterparty effects have to be taken into account.

Let $A$ denote the protection buyer, $B$ denote the protection seller and $C$ denote the reference entity. The binomial default rule considers only two possible states: default or survival. Therefore, the default indicator $Y_j$ for firm $j$ ($j = A$ or $B$ or $C$) follows a Bernoulli distribution, which takes value 1 with default probability $q_j$, and value 0 with survival probability $p_j$. The marginal default distributions can be determined by the reduced-form models. The joint distributions of a multivariate Bernoulli variable can be easily obtained via the marginal distributions by introducing extra correlations. The joint probability representations of a trivariate Bernoulli distribution (see Teugels (1990)) are given by

$$p_{000} := P(Y_A = 0, Y_B = 0, Y_C = 0) = p_A p_B p_C + p_C \sigma_{AB} + p_B \sigma_{AC} + p_A \sigma_{BC} - \theta_{ABC} \tag{3a}$$

$$p_{100} := P(Y_A = 1, Y_B = 0, Y_C = 0) = q_A p_B p_C - p_C \sigma_{AB} - p_B \sigma_{AC} + q_A \sigma_{BC} + \theta_{ABC} \tag{3b}$$

$$p_{010} := P(Y_A = 0, Y_B = 1, Y_C = 0) = p_A q_B p_C - p_C \sigma_{AB} + q_B \sigma_{AC} - p_A \sigma_{BC} + \theta_{ABC} \tag{3c}$$

$$p_{001} := P(Y_A = 0, Y_B = 0, Y_C = 1) = p_A p_B q_C + q_C \sigma_{AB} - p_B \sigma_{AC} - p_A \sigma_{BC} + \theta_{ABC} \tag{3d}$$

$$p_{110} := P(Y_A = 1, Y_B = 1, Y_C = 0) = q_A q_B p_C + p_C \sigma_{AB} - q_B \sigma_{AC} - q_A \sigma_{BC} - \theta_{ABC} \tag{3e}$$

$$p_{101} := P(Y_A = 1, Y_B = 0, Y_C = 1) = q_A p_B q_C - q_C \sigma_{AB} + p_B \sigma_{AC} - q_A \sigma_{BC} - \theta_{ABC} \tag{3f}$$

$$p_{011} := P(Y_A = 0, Y_B = 1, Y_C = 1) = p_A q_B q_C - q_C \sigma_{AB} - q_B \sigma_{AC} + p_A \sigma_{BC} - \theta_{ABC} \tag{3g}$$

$$p_{111} := P(Y_A = 1, Y_B = 1, Y_C = 1) = q_A q_B q_C + q_C \sigma_{AB} + q_B \sigma_{AC} + q_A \sigma_{BC} + \theta_{ABC} \tag{3h}$$

where

$$\theta_{ABC} := E\big((Y_A - q_A)(Y_B - q_B)(Y_C - q_C)\big) \tag{3i}$$

Equation (3) tells us that the joint probability distribution of three defaultable parties depends not only on the bivariate statistical relationships of all pair-wise combinations (e.g., $\sigma_{ij}$) but also on the trivariate statistical relationship (e.g., $\theta_{ABC}$). $\theta_{ABC}$ was first defined by Deardorff (1982) as *comvariance*, who use it to correlate three random variables that are the value of commodity net imports/exports, factor intensity, and factor abundance in international trading.

We introduce the concept of *comvariance* into credit risk modeling arena to exploit any statistical relationship among multiple random variables. Furthermore, we define a new statistic, *comrelation*, as a scaled version of comvariance (just like correlation is a scaled version of covariance) as follows:

***Definition 1****: For three random variables* $X_A$, $X_B$, *and* $X_C$, *let* $\mu_A$, $\mu_B$, *and* $\mu_C$ *denote the means of* $X_A$, $X_B$, *and* $X_C$. *The comrelation of* $X_A$, $X_B$, *and* $X_C$ *is defined by*

$$\zeta_{ABC} = \frac{E[(X_A - \mu_A)(X_B - \mu_B)(X_C - \mu_C)]}{\sqrt[3]{E|X_A - \mu_A|^3 \times E|X_B - \mu_B|^3 \times E|X_C - \mu_C|^3}} \tag{4}$$

According to the Holder inequality, we have

$$\left|E\left((X_A-\mu_A)(X_B-\mu_B)(X_C-\mu_C)\right)\right| \le E\left|(X_A-\mu_A)(X_B-\mu_B)(X_C-\mu_C)\right| \\ \le \sqrt[3]{E|X_A-\mu_A|^3 \times E|X_B-\mu_B|^3 \times E|X_C-\mu_C|^3} \tag{5}$$

Obviously, the comrelation is in the range of [-1, 1]. Given the comrelation, Equation (3i) can be rewritten as

$$\theta_{ABC} := E\left((Y_A-q_A)(Y_B-q_B)(X_C-q_C)\right) = \zeta_{ABC}\sqrt[3]{E|Y_A-q_A|^3 \times E|Y_B-q_B|^3 \times E|Y_C-q_C|^3} \\ = \zeta_{ABC}\sqrt[3]{\left|p_A q_A(p_A^2+q_A^2)p_B q_B(p_B^2+q_B^2)p_C q_C(p_C^2+q_C^2)\right|} \tag{6}$$

where $E(Y_j)=q_j$ and $E|Y_j-q_j|^3 = p_j q_j(p_j^2+q_j^2)$, *j=A, B, or C.*

If we have a series of *n* measurements of $X_A$, $X_B$, *and* $X_C$ written as $x_{Ai}$, $x_{Bi}$ and $x_{Ci}$ where *i* = 1,2,…,*n*, the sample *comrelation coefficient* can be obtained as:

$$\zeta_{ABC} = \frac{\sum_{i=1}^{n}(x_{Ai}-\mu_A)(x_{Bi}-\mu_B)(x_{Ci}-\mu_C)}{\sqrt[3]{\sum_{i=1}^{n}|x_{Ai}-\mu_A|^3 \times \sum_{i=1}^{n}|x_{Bi}-\mu_B|^3 \times \sum_{i=1}^{n}|x_{Ci}-\mu_C|^3}} \tag{7}$$

More generally, we define the *comrelation* in the context of *n* random variables as

***Definition 2****: For n random variables* $X_1$, $X_2$,..., $X_n$*, let* $\mu_i$ *denote the mean of* $X_i$ *where i=1,..,n. The comrelation of* $X_1$, $X_2$,..., $X_n$ *is defined as*

$$\zeta_{12..n} = \frac{E\left[(X_1-\mu_1)(X_2-\mu_2)\cdots(X_n-\mu_n)\right]}{\sqrt[n]{E|X_1-\mu_1|^n \times E|X_2-\mu_2|^n \cdots \times E|X_n-\mu_n|^n}} \tag{8}$$

*Correlation is just a specific case of comrelation where n = 2.* Again, the comrelation $\zeta_{12..n}$ is in the range of [-1, 1] according to the Holder inequality.

### 2.1 Risky valuation without collateralization

Recovery assumptions are important for pricing credit derivatives. If the reference entity under a CDS contract defaults, the best assumption, as pointed out by J. P. Morgan (1999), is that the recovered value equals the recovery rate times the face value plus accrued interest[2]. In other words, the recovery of par value assumption is a better fit upon the default of the reference entity, whereas the recovery of market value assumption is a more suitable choice in the event of a counterparty default[3].

[2] In the market, there is an average accrual premium assumption, i.e., the average accrued premium is half the full premium due to be paid at the end of the premium.

Let valuation date be $t$. Suppose that a CDS has $m$ scheduled payments represented as $X_i = -sN\delta(T_{i-1},T_i)$ with payment dates $T_1,\ldots,T_m$ where *i=1,,,,m,* $\delta(T_{i-1},T_i)$ denotes the accrual factor for period $(T_{i-1},T_i)$, $N$ denotes the notional/principal, and $s$ denotes the CDS premium. Party $A$ pays the premium/fee to party $B$ if reference entity $C$ does not default. In return, party $B$ agrees to pay the protection amount to party $A$ if reference entity $C$ defaults before the maturity. We have the following proposition.

***Proposition 1:*** *The value of the CDS is given by*

$$V(t) = \sum_{i=1}^{m} E\left[\left(\prod_{j=0}^{i-1} O(T_j,T_{j+1})\right) X_i \middle| \mathcal{F}_t\right] + \sum_{i=1}^{m} E\left[\left(\prod_{j=0}^{i-2} O(T_j,T_{j+1})\right) \Omega(T_{i-1},T_i) R(T_{i-1},T_i) \middle| \mathcal{F}_t\right] \tag{9a}$$

*where* $t = T_0$ *and*

$$O(T_j,T_{j+1}) = 1_{\left(V(T_{j+1})+X_{j+1}\right)\geq 0}\,\phi_B(T_j,T_{j+1}) + 1_{\left(V(T_{j+1})+X_{j+1}\right)<0}\,\phi_A(T_j,T_{j+1}) \tag{9b}$$

$$\begin{aligned}
\phi_A(T_j,T_{j+1}) = \Big\{ & p_A(T_j,T_{j+1}) p_B(T_j,T_{j+1}) p_C(T_j,T_{j+1}) + q_A(T_j,T_{j+1}) p_B(T_j,T_{j+1}) p_C(T_j,T_{j+1}) \varphi_A(T_{j+1}) \\
& + p_A(T_j,T_{j+1}) q_B(T_j,T_{j+1}) p_C(T_j,T_{j+1}) \overline{\varphi}_A(T_{j+1}) + q_A(T_j,T_{j+1}) q_B(T_j,T_{j+1}) p_C(T_j,T_{j+1}) \varphi_{AB}(T_{j+1}) \\
& + p_C(T_j,T_{j+1}) \sigma_{AB}(T_j,T_{j+1}) \left(1 - \varphi_A(T_{j+1}) - \overline{\varphi}_A(T_{j+1}) + \varphi_{AB}(T_{j+1})\right) \\
& + \sigma_{AC}(T_j,T_{j+1}) \left[p_B(T_j,T_{j+1})\left(1-\varphi_A(T_{j+1})\right) + q_B(T_j,T_{j+1})\left(\overline{\varphi}_A(T_{j+1}) - \varphi_{AB}(T_{j+1})\right)\right] \\
& + \sigma_{BC}(T_j,T_{j+1}) \left[p_A(T_j,T_{j+1})\left(1-\overline{\varphi}_A(T_{j+1})\right) + q_A(T_j,T_{j+1})\left(\varphi_A(T_{j+1}) - \varphi_{AB}(T_{j+1})\right)\right] \\
& + \theta_{ABC}(T_j,T_{j+1}) \left(-1 + \overline{\varphi}_A(T_{j+1}) - \varphi_{AB}(T_{j+1}) + \varphi_A(T_{j+1})\right) \Big\} D(T_j,T_{j+1})
\end{aligned} \tag{9c}$$

$$\begin{aligned}
\phi_B(T_j,T_{j+1}) = \Big\{ & p_A(T_j,T_{j+1}) p_B(T_J,T_{j+1}) p_C(T_j,T_{j+1}) + q_A(T_j,T_{j+1}) p_B(T_j,T_{j+1}) p_C(T_j,T_{j+1}) \overline{\varphi}_B(T_{j+1}) \\
& + p_A(T_j,T_{j+1}) q_B(T_j,T_{j+1}) p_C(T_j,T_{j+1}) \varphi_B(T_{j+1}) + q_A(T_j,T_{j+1}) q_B(T_j,T_{j+1}) p_C(T_j,T_{j+1}) \varphi_{AB}(T_{j+1}) \\
& + p_C(T_j,T_{j+1}) \sigma_{AB}(T_j,T_{j+1}) \left(1 - \varphi_B(T_{j+1}) - \overline{\varphi}_B(T_{j+1}) + \varphi_{AB}(T_{j+1})\right) \\
& + \sigma_{AC}(T_j,T_{j+1}) \left[p_B(T_j,T_{j+1})\left(1-\overline{\varphi}_B(T_{j+1})\right) + q_B(T_j,T_{j+1})\left(\varphi_B(T_{j+1}) - \varphi_{AB}(T_{j+1})\right)\right] \\
& + \sigma_{BC}(T_j,T_{j+1}) \left[p_A(T_j,T_{j+1})\left(1-\varphi_B(T_j)\right) + q_A(T_j,T_{j+1})\left(\overline{\varphi}_B(T_{j+1}) - \varphi_{AB}(T_{j+1})\right)\right] \\
& + \theta(T_j,T_{j+1}) \left(-1 + \overline{\varphi}_B(T_{j+1}) - \varphi_{AB}(T_{j+1}) + \varphi_B(T_{j+1})\right) \Big\} D(T_j,T_{j+1})
\end{aligned} \tag{9d}$$

---

[3] Three different recovery models exist in the literature. The default payoff is either i) a fraction of par (Madan and Unal (1998)), ii) a fraction of an equivalent default-free bond (Jarrow and Turnbull (1995)), or iii) a fraction of market value (Duffie and Singleton (1999)).

$$
\begin{aligned}
\Omega(T_j,T_{j+1}) = \{ & p_A(T_j,T_{j+1})p_B(T_j,T_{j+1})q_C(T_j,T_{j+1}) + q_A(T_j,T_{j+1})p_B(T_j,T_{j+1})q_C(T_j,T_{j+1})\overline{\varphi}_B(T_{j+1}) \\
& + p_A(T_j,T_{j+1})q_B(T_j,T_{j+1})q_C(T_j,T_{j+1})\varphi_B(T_{j+1}) + q_A(T_j,T_{j+1})q_B(T_j,T_{j+1})q_C(T_j,T_{j+1})\varphi_{AB}(T_{j+1}) \\
& + q_C(T_j,T_{j+1})\sigma_{AB}(T_j,T_{j+1})\left(1-\varphi_B(T_{j+1})-\overline{\varphi}_B(T_{j+1})+\varphi_{AB}(T_{j+1})\right) \\
& - \sigma_{AC}(T_j,T_{j+1})\left[p_B(T_j,T_{j+1})\left(1-\overline{\varphi}_B(T_{j+1})\right)+q_B(T_j,T_{j+1})\left(\varphi_B(T_{j+1})-\varphi_{AB}(T_{j+1})\right)\right] \\
& - \sigma_{BC}(T_j,T_{j+1})\left[p_A(T_j,T_{j+1})\left(1-\varphi_B(T_{j+1})\right)+q_A(T_j,T_{j+1})\left(\overline{\varphi}_B(T_{j+1})-\varphi_{AB}(T_{j+1})\right)\right] \\
& + \theta_{ABC}(T_j,T_{j+1})\left(1-\overline{\varphi}_B(T_{j+1})+\varphi_{AB}(T_{j+1})-\varphi_B(T_{j+1})\right)\}D(T_j,T_{j+1})
\end{aligned}
\tag{9e}
$$

where $R(T_j,T_{j+1}) = \left(N\left(1-\varphi_C(T_{j+1})\right)-\alpha(T_j,T_{j+1})\right)$, $\alpha(T_j,T_{j+1}) = sN\delta(T_S,T)/2$, and $X_i = -sN\delta(T_j,T_{j+1})$.

Proof: See the Appendix.

We may think of $O(t,T)$ as the risk-adjusted discount factor for the premium and $\Omega(t,T)$ as the risk-adjusted discount factor for the default payment. Proposition 1 says that the pricing process of a multiple-payment instrument has a backward nature since there is no way of knowing which risk-adjusted discounting rate should be used without knowledge of the future value. Only on the maturity date, the value of an instrument and the decision strategy are clear. Therefore, the evaluation must be done in a backward fashion, working from the final payment date towards the present. This type of valuation process is referred to as backward induction.

Proposition 1 provides a general form for pricing a CDS. Applying it to a particular situation in which we assume that counterparties *A and B* are default-free, i.e., $p_j = 1$, $q_j = 0$, $\rho_{kl} = 0$, and $\varsigma_{ABC} = 0$, where *j=A or B* and *k, l=A, B, or C,* we derive the following corollary.

***Corollary 1:*** *If counterparties A and B are default-free, the value of the CDS is given by*

$$
\begin{aligned}
V(t) &= \sum_{i=1}^{m} E\left[\left(\prod_{j=0}^{i-1} O(T_j,T_{j+1})\right)X_i \middle| \mathcal{F}_t\right] + \sum_{i=1}^{m} E\left[\left(\prod_{j=0}^{i-2} O(T_j,T_{j+1})\right)\Omega(T_{i-1},T_i)R(T_{i-1},T_i) \middle| \mathcal{F}_t\right] \\
&= \sum_{i=1}^{m} E\left[D(t,T_i)p_C(t,T_i)X_i \middle| \mathcal{F}_t\right] + \sum_{i=1}^{m} E\left[D(t,T_i)p_C(t,T_{i-1})q_C(T_{i-1},T_i)R(T_{i-1},T_i) \middle| \mathcal{F}_t\right]
\end{aligned}
\tag{10}
$$

*where* $O(T_{i-1},T_i) = D(T_{i-1},T_i)p_C(T_{i-1},T_i)$; $\Omega(T_{i-1},T_i) = D(T_{i-1},T_i)q_C(T_{i-1},T_i)$.

The proof of this corollary becomes straightforward according to Proposition 1 by setting $\rho_{kl}$ =0, $\varphi_{AB} = 0$, $\varsigma_{ABC} = 0$, $p_j = 1$, $q_j = 0$, $p_C(t,T_i) = \prod_{g=0}^{i-1} p(T_g,T_{g+1})$, and $D(t,T_i) = \prod_{g=0}^{i-1} D(T_g,T_{g+1})$.

If we further assume that the discount factor and the default probability of the reference entity are uncorrelated and the recovery rate $\varphi_C$ is constant, we have

***Corollary 2:*** *Assume that i) counterparties A and B are default-free, ii) the discount factor and the default probability of the reference entity are uncorrelated; iii) the recovery rate* $\varphi_C$ *is constant; the value of the CDS is given by*

$$V(t)=\sum_{i=1}^{m}P(t,T_i)\bar{p}_C(t,T_{i-1})\bar{q}_C(T_{i-1},T_i)\left(N\left(1-\varphi_C\right)-\alpha(T_{i-1},T_i)\right)-\sum_{i=1}^{m}P(t,T_i)\bar{p}_c(t,T_i)sN\delta(T_{i-1},T_i) \tag{11}$$

*where* $P(t,T_i)=E\left[D(t,T_i)\middle|\mathcal{F}_t\right]$ *denotes the bond price,* $\bar{p}_c(t,T_i)=E\left[p_c(t,T_i)\middle|\mathcal{F}_t\right]$, $\bar{q}_c(t,T_i)=1-\bar{p}_c(t,T_i)$, $\bar{p}(t,T_{i-1})\bar{q}(T_{i-1},T_i)=\bar{p}(t,T_{i-1})-\bar{p}(t,T_i)$.

This corollary is easily proved according to Corollary 1 by setting $E\left[XY\middle|\mathcal{F}_t\right]=E\left[X\middle|\mathcal{F}_t\right]E\left[Y\middle|\mathcal{F}_t\right]$ when $X$ and $Y$ are uncorrelated. *Corollary 2 is the formula for pricing CDS in the market*.

Our methodology can be extended to the cases where the number of parties $n\geq 4$. A generating function for the (probability) joint distribution (see details in Teugels (1990)) of *n*-variate Bernoulli can be expressed as

$$p^{(n)}=\begin{bmatrix}p_n & -1\\ q_n & 1\end{bmatrix}\otimes\begin{bmatrix}p_{n-1} & -1\\ q_{n-1} & 1\end{bmatrix}\otimes\cdots\otimes\begin{bmatrix}p_1 & -1\\ q_1 & 1\end{bmatrix}\sigma^{(n)} \tag{12}$$

where $\otimes$ denotes the Kronecker product; $p^{(n)}=\left\{p_k^{(n)}\right\}$ and $\sigma^{(n)}=\left\{\sigma_k^{(n)}\right\}$ are vectors containing $2^n$ components: $p_k^{(n)}=p_{k_1,k_2,\ldots k_n}$, $k=1+\sum_{i=1}^{n}k_i 2^{i-1}$, $k_i\in\{0,1\}$; $\sigma_k^{(n)}=\sigma_{k_1,k_2,\ldots k_n}=E\left(\prod_{i=1}^{n}\left(Y_i-q_i\right)^{k_i}\right)$.

## 2.2 Risky valuation with collateralization

Collateralization is the most important and widely used technique in practice to mitigate credit risk. The posting of collateral is regulated by the Credit Support Annex (CSA) that specifies a variety of terms including the threshold, the independent amount, and the minimum transfer amount (MTA), etc. The threshold is the unsecured credit exposure that a party is willing to bear. The minimum transfer amount is the smallest amount of collateral that can be transferred. The independent amount plays the same role as the initial margin (or haircuts).

In a typical collateral procedure, a financial instrument is periodically marked-to-market and the collateral is adjusted to reflect changes in value. The collateral is called as soon as the mark-to-market (MTM) value rises above the given collateral threshold, or more precisely, above the threshold amount plus the minimum transfer amount. Thus, the collateral amount posted at time $t$ is given by

$$C(t) = \begin{cases} V(t) - H(t) & if\ V(t) > H(t) \\ 0 & otherwise \end{cases} \tag{13}$$

where $H(t)$ is the collateral threshold. In particular, $H(t) = 0$ corresponds to full-collateralization[4]; $H > 0$ represents partial/under-collateralization; and $H < 0$ is associated with over-collateralization. Full collateralization becomes increasingly popular at the transaction level. In this paper, we focus on full collateralization only, i.e., $C(t) = V(t)$.

The main role of collateral should be viewed as an improved recovery in the event of a counterparty default. According to Bankruptcy law, if there has been no default, the collateral is returned to the collateral giver by the collateral taker. If a default occurs, the collateral taker possesses the collateral. In other words, collateral does not affect the survival payment; instead, it takes effect on the default payment only.

According to the ISDA (2013), almost all CDSs are fully collateralized. Many people believe that full collateralization can eliminate counterparty risk completely for CDS.

Collateral posting regimes are originally designed and utilized for bilateral risk products, e.g., IRS, but there are many reasons to be concerned about the success of collateral posting in offsetting the risks of CDS contracts. First, the values of CDS contracts tend to move very suddenly with big jumps, whereas the price movements of IRS contracts are far smoother and less volatile than CDS prices. Second, CDS spreads can widen very rapidly. The amount of collateral that one party is required to provide at short notice may, in some cases, be close to the notional amount of the CDS and may therefore exceed that party's short-term liquidity capacity, thereby triggering a liquidity crisis. Third, CDS contracts have many more risk factors than IRS contracts.

[4] There are three types of collateralization: Full-collateralization is a process where the posting of collateral is equal to the current MTM value. Partial/under-collateralization is a process where the posting of collateral is less than the current MTM value. Over-collateralization is a process where the posting of collateral is greater than the current MTM value.

We assume that a CDS is fully collateralized, i.e., the posting of collateral is equal to the amount of the current *MTM* value: $C(t) = V(t)$. For a discrete one-period (*t*, *u*) economy, there are several possible states at time *u*: i) *A*, *B*, and *C* survive with probability $p_{000}$. The instrument value is equal to the market value $V(u)$; ii) *A* and *B* survive, but *C* defaults with probability $p_{001}$. The instrument value is the default payment $R(u)$; iii) For the remaining cases, either or both counterparties *A* and *B* default. The instrument value is the future value of the collateral $V(t)/D(t,u)$ (Here we consider the time value of money only). The value of the collateralized instrument at time *t* is the discounted expectation of all the payoffs and is given by

$$\begin{aligned} V(t) = E\{&D(t,u)[p_{000}(t,u)V(u) + p_{001}(t,u)R(u) \\ &+ (p_{100}(t,u) + p_{010}(t,u) + p_{110}(t,u) + p_{101}(t,u) + p_{011}(t,u) + p_{111}(t,u))V(t)/D(t,u)] | \mathcal{F}_t\} \\ = E\{&[D(t,u)(p_{000}(t,u)V(s) + p_{001}(t,u)R(u)) + (1 - p_{000}(t,u) - p_{001}(t,u))V(t)] | \mathcal{F}_t\} \end{aligned} \tag{14a}$$

or

$$\begin{aligned} &E[(p_A(t,u)p_B(t,u) + \sigma_{AB}(t,u)) | \mathcal{F}_t]V(t) \\ &\quad = E\{[D(t,u)(p_C(t,u)V(u) + q_C(t,u)R(u))(p_A(t,u)p_B(t,u) + \sigma_{AB}(t,u)) \\ &\quad\quad + D(t,u)(V(u) - R(u))(p_B(t,u)\sigma_{AC}(t,u) + p_A(t,u)\sigma_{BC}(t,u) - \theta_{ABC}(t,u))] | \mathcal{F}_t\} \end{aligned} \tag{14b}$$

If we assume that $(p_A(t,u)p_B(t,u) + \sigma_{AB}(t,u))$ and $D(t,u)(p_C(t,u)V(u) + q_C(t,u)R(u))$ are uncorrelated, we have

$$V(t) = V^F(t) + \xi_{ABC}(t,u)/\psi_{AB}(t,u) \tag{15a}$$

where

$$V^F(t) = E\{D(t,u)[p_C(t,u)V(u) + q_C(t,u)R(u)] | \mathcal{F}_t\} \tag{15b}$$

$$\psi_{AB}(t,u) = E\{[p_A(t,u)p_B(t,u) + \sigma_{AB}(t,u)] | \mathcal{F}_t\} \tag{15c}$$

$$\xi_{ABC}(t,u) = E\{[D(t,u)(p_B(t,u)\sigma_{AC}(t,u) + p_A(t,u)\sigma_{BC}(t,u) - \theta_{ABC}(t,u))(V(u) - R(u))] | \mathcal{F}_t\} \tag{15d}$$

The first term $V^F(t)$ in equation (15) is the counterparty-risk-free value of the CDS and the second term is the exposure left over under full collateralization, which can be substantial.

***Proposition 2****: If a CDS is fully collateralized, the risky value of the CDS is NOT equal to the counterparty-risk-free value, as shown in equation (15).*

Proposition 2 or equation (15) provides a theoretical explanation for the failure of full collateralization in the CDS market. It tells us that under full collateralization the risky value is in general not equal to the counterparty-risk-free value except in one of the following situations: i) the market value is equal to the default payment, i.e., $V(u) = R(u)$; ii) firms *A*, *B*, and *C* have independent credit risks, i.e., $\rho_{ij}=0$ and $\varsigma_{ABC} = 0$; or iii) the following relationship holds $p_B\sigma_{AC} + p_A\sigma_{BC} = \theta_{ABC}$.

## 3 Numerical Results

Our goal in this section is to study the quantitative relationship between CDS premia and the credit quality of counterparties and reference entities, including the default correlations and comrelations.

In our study, we choose a new 5-year CDS with a quarterly payment frequency. Two counterparties are denoted as *A* and *B*. Counterparty *A* buys a protection from counterparty *B*. All calculations are from the perspective of party *A*. By definition, a breakeven CDS spread is a premium that makes the market value of a given CDS at inception zero.

The current (spot) market data are shown in Table 1. Assume that the reference entity *C* has an "A+200bps" credit quality throughout this subsection. The 5-year counterparty-risk-free CDS premium is 0.027 (equals the 5-year 'A' rated CDS spread in Table 1 plus 200 basis points).

Since the payoffs of a CDS are mainly determined by credit events, we need to characterize the evolution of the hazard rates. Here we choose the *Cox-Ingersoll-Ross* (CIR) model. The CIR process has been widely used in the literature of credit risk and is given by

$$dh_t = a(b - h_t)dt + \sigma\sqrt{h_t}\,dW_t \tag{16}$$

where $a$ denotes the mean reversion speed, *b* denotes the long-term mean, and $\sigma$ denotes the volatility.

**Table 1: Current/spot market data**

This table displays the current (spot) market data used for all calculations in this paper, including the term structure of continuously compounded interest rates, the term structure of A-rated breakeven CDS spreads, and the curve of at-the-money caplet volatilities.

| Term (days) | 31 | 91 | 182 | 365 | 548 | 730 | 1095 | 1825 | 2555 | 3650 | 5475 |
|---|---|---|---|---|---|---|---|---|---|---|---|
| **Interest Rate** | 0.0028 | 0.0027 | 0.0029 | 0.0043 | 0.0071 | 0.0102 | 0.016 | 0.0249 | 0.0306 | 0.0355 | 0.0405 |
| **Credit Spread** | 0.0042 | 0.0042 | 0.0042 | 0.0045 | 0.0049 | 0.0052 | 0.0058 | 0.007 | 0.0079 | 0.0091 | 0.0106 |
| **Caplet Volatility** | 0.3267 | 0.331 | 0.3376 | 0.3509 | 0.3641 | 0.3773 | 0.308 | 0.2473 | 0.2141 | 0.1678 | 0.1634 |

**Table 2: Risk-neutral parameters for CIR model**

This table presents the risk-neutral parameters that are calibrated to the current market shown in Table 1. ‘A+100bps’ represents a ‘100 basis points’ parallel shift in the A-rated CDS spreads.

| Credit Quality | A | A+100bps | A+200bps | A+300bps |
|---|---|---|---|---|
| **Long-Term Mean $a$** | 0.035 | 0.056 | 0.077 | 0.099 |
| **Mean Reverting Speed *b*** | 0.14 | 0.18 | 0.25 | 0.36 |
| **Volatility $\sigma$** | 0.022 | 0.028 | 0.039 | 0.056 |

The calibrated parameters are shown in table 2. We assume that interest rates are deterministic and select the regression-based Monte-Carlo simulation (see Longstaff and Schwartz (2001)) to perform risky valuation.

We first assume that counterparties *A*, *B*, and reference entity *C* have independent default risks, i.e., $\rho_{AB} = \rho_{AC} = \rho_{BC} = \varphi_{AB} = \zeta_{ABC} = 0$, and examine the following cases: i) *B* is risk-free and *A* is risky; and ii) *A* is risk-free and *B* is risky. We simulate the hazard rates using the CIR model and then determine the appropriate discount factors according to Proposition 1. Finally we calculate the prices via the regression-based Monte-Carlo method. The results are shown in Table 3 and 4.

**Table 3: Impact of the credit quality of the protection buyer on CDS premia**

This table shows how the CDS premium increases as the credit quality of party *A* decreases. The 1st data column represents the counterparty-risk-free results. For the remaining columns, we assume that party *B* is risk-free and party *A* is risky. 'A+100bps' represents a '100 basis points' parallel shift in the A-rated CDS spreads. The results in the row 'Difference from Risk-Free' = current CDS premium – counterparty-risk-free CDS premium.

| **Credit Quality** | **Party *A*** | - | A | A+100bps | A+200bps | A+300bps |
|---|---|---|---|---|---|---|
| | **Party *B*** | - | - | - | - | - |
| **CDS premium** | | 0.027 | 0.02703 | 0.02708 | 0.02713 | 0.02717 |
| **Difference from Risk-Free** | | 0 | 0.003% | 0.008% | 0.013% | 0.017% |

**Table 4: Impact of the credit quality of the protection seller on CDS premia**

This table shows the decrease in the CDS premium with the credit quality of party *B*. The 1st data column represents the counterparty-risk-free results. For the remaining columns, we assume that party *A* is risk-free and party *B* is risky. 'A+100bps' represents a '100 basis points' parallel shift in the A-rated CDS spreads. The results in the row 'Difference from Risk-Free' = current CDS premium – counterparty-risk-free CDS premium.

| **Credit Quality** | **Party A** | - | - | - | - | - |
|---|---|---|---|---|---|---|
| | **Party B** | - | A | A+100bps | A+200bps | A+300bps |
| **CDS premium** | | 0.027 | 0.02695 | 0.02687 | 0.0268 | 0.02672 |
| **Difference from Risk-Free** | | 0.00% | -0.005% | -0.013% | -0.020% | -0.028% |

From table 3 and 4, we find that a credit spread of about 100 basis points maps into a CDS premium of about 0.4 basis points for counterparty *A* and about -0.7 basis points for counterparty *B*. The credit impact on the CDS premia is approximately linear. As would be expected, i) the dealer's credit quality has a larger impact on CDS premia than the investor's credit quality; ii) the higher the investor's credit risk, the higher the premium that the dealer charges; iii) the higher the dealer's credit risk, the lower the premium that the dealer asks. Without considering default correlations and comrelations, we find that, in general, the impact of counterparty risk on CDS premia is relatively small. This is in line with the empirical findings of Arora, Gandhi, and Longstaff (2009).

**Figure 1: Impact of default correlations and comrelation on CDS premia**

Each curve in this figure illustrates how CDS premium changes as default correlations and comrelation move from -1 to 1. For instance, the curve 'cor_BC' represents the sensitivity of the CDS premium to changes in the correlation $\rho_{BC}$ when $\rho_{AB} = \rho_{AC} = \zeta_{ABC} = 0$.

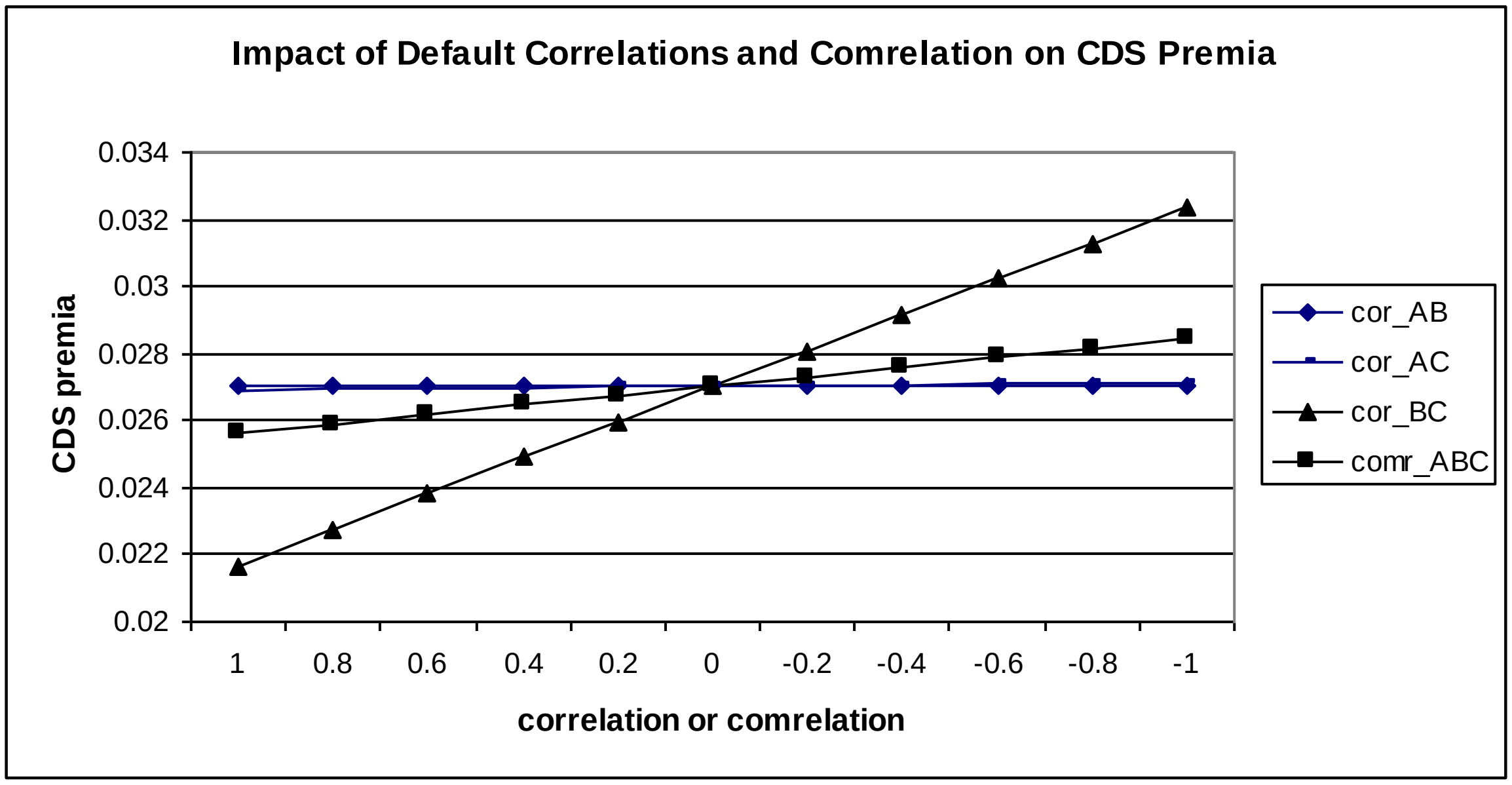

Next, we study the sensitivity of CDS premia to changes in the joint credit quality of associated parties. Sensitivity analysis is a very popular way in finance to find out how the value and risk of an instrument/portfolio changes if risk factors change. One of the simplest and most common approaches involves changing one factor at a time to see what effect this produces on the output. We are going to examine the impacts of the default correlations $\rho_{AB}$, $\rho_{AC}$, $\rho_{BC}$, and the comrelation $\zeta_{ABC}$ separately. Assume that party *A* has an 'A+100bps' credit quality and party *B* has an 'A' credit quality. The 5-year risky CDS premium is calculated as 0.02703.

Assume $\varphi_{AB}$=0.5. The impact diagrams of the default correlations and comrelation are shown in Figure 1. From this graph, we can draw the following conclusions: First, the CDS premium and the default correlations/comrelation have a negative relation. Intuitively, a protection seller who is positively correlated with the reference entity (a wrong way risk) should charge a lower premium for selling credit protection. Next, the impacts of the default correlations and comrelation are approximately linear. Finally, the sensitivity slopes of the CDS premium to the default correlations and comrelation are -0.06 to $\rho_{AB}$; -0.09 to $\rho_{AC}$; -53 to $\rho_{BC}$; and -14 to $\zeta_{ABC}$. Slope measures the rate of change in the premium as a result of a change in the default dependence. For instance, a slope of -53 implies that the CDS premium would have to decrease by 53 basis points when a default correlation/comrelation changes from 0 to 1.

As the absolute value of the slope increases, so does the sensitivity. The results illustrate that $\rho_{BC}$ has the largest effect on CDS premia. The second biggest one is $\zeta_{ABC}$. The impacts of $\rho_{AB}$ and $\rho_{AC}$ are very small. In particular, the effect of the comrelation is substantial and has never been studies before. A natural intuition to have on CDS is that the party buying default protection should worry about the default correlations and comrelation.

## 4 Conclusion

This article presents a new valuation framework for pricing financial instruments subject to credit risk. In particular, we focus on modeling default relationships.

To capture the default relationships among more than two defaultable entities, we introduce a new statistic: *comrelation*, an analogue to correlation for multiple variables, to exploit any multivariate statistical relationship. Our research shows that accounting for default correlations and comrelations becomes important, especially under market stress. The existing valuation models in the credit derivatives market, which take into account only pair-wise default correlations, may underestimate credit risk and may be inappropriate.

We study the sensitivity of the price of a defaultable instrument to changes in the joint credit quality of the parties. For instance, our analysis shows that the effect of default dependence on CDS premia from large to small is the correlation between the protection seller and the reference entity, the comrelation, the correlation between the protection buyer and the reference entity, and the correlation between the protection buyer and the protection seller.

The model shows that a fully collateralized CDS is not equivalent to a risk-free one. Therefore, we conclude that collateralization designed to mitigate counterparty risk works well for financial instruments subject to bilateral credit risk, but fails for ones subject to multilateral credit risk.

## Appendix

**Proof of Proposition 1.** Let $t = T_0$. On the first payment date $T_1$, let $V(T_1)$ denote the market value of the CDS excluding the current cash flow $X_1$. There are a total of eight ( $2^3 = 8$ ) possible states shown in Table A_1.

**Table A_1. Payoffs of a trilaterally defaultable CDS**

This table shows all possible payoffs at time $T_1$. In the case of $V(T_1) + X_1 \geq 0$ where $V(T_1)$ is the market value excluding the current cash flow $X_1$, there are a total of eight ( $2^3 = 8$ ) possible states: i) *A*, *B,* and *C* survive with probability $p_{000}$. The instrument value equals the market value: $V(T_1) + X_1$. ii) *A* defaults, but *B* and *C* survive with probability $p_{100}$. The instrument value is a fraction of the market value: $\overline{\varphi}_B(T_1)\left(V(T_1) + X_1\right)$ where $\overline{\varphi}_B$ represents the non-default recovery rate of party *B*[5]. $\overline{\varphi}_B$ =0 represents the one-way settlement rule, while $\bar{\varphi}_B$ =1 represents the two-way settlement rule. iii) *A* and *C* survive, but *B* defaults with probability $p_{010}$. The instrument value is given by $\varphi_B(T_1)(V(T_1)+X_1)$ where $\varphi_B$ represents the default recovery rate of defaulting party *B*. iv) *A* and *B* survive, but *C* defaults with probability $p_{001}$. The instrument value is the default payment: $R(T_0,T_1)$. v) *A* and *B* default, but *C* survives with probability $p_{110}$. The instrument value is given by $\varphi_{AB}(T_1)(V(T_1)+X_1)$ where $\varphi_{AB}$ denotes the joint recovery rate when both parties *A* and *B* default simultaneously. vi) *A* and *C* default, but *B* survives with probability $p_{101}$. The instrument value is a fraction of the default payment: $\bar{\varphi}_B(T)R(T_0,T_1)$. vii) *B* and *C* default, but *A* survives with probability $p_{011}$, The instrument value is given by $\varphi_B(T)R(T_0,T_1)$. viii) *A*, *B*, and *C* default with probability $p_{111}$. The instrument value is given by $\varphi_{AB}(T)R(T_0,T_1)$. A similar logic applies to the case of $V(T_1)+X_1<0$.

[5] There are two default settlement rules in the market. The *one-way payment rule* was specified by the early ISDA master agreement. The non-defaulting party is not obligated to compensate the defaulting party if the remaining market value of the instrument is positive for the defaulting party. The *two-way*

| **Status** | **Probability** | **Payoff if** $V(T_1)+X_1\geq 0$ | **Payoff if** $V(T_1)+X_1<0$ |
|---|---|---|---|
| $Y_A=0,Y_B=0,Y_C=0$ | $p_{000}$ | $V(T_1)+X_1$ | $V(T_1)+X_1$ |
| $Y_A=1,Y_B=0,Y_C=0$ | $p_{100}$ | $\bar{\varphi}_B(T_1)(V(T_1)+X_1)$ | $\varphi_A(T_1)(V(T_1)+X_1)$ |
| $Y_A=0,Y_B=1,Y_C=0$ | $p_{010}$ | $\varphi_B(T_1)(V(T_1)+X_1)$ | $\bar{\varphi}_A(T_1)(V(T_1)+X_1)$ |
| $Y_A=0,Y_B=0,Y_C=1$ | $p_{001}$ | $R(T_0,T_1)$ | $R(T_0,T_1)$ |
| $Y_A=1,Y_B=1,Y_C=0$ | $p_{110}$ | $\varphi_{AB}(T_1)(V(T_1)+X_1)$ | $\varphi_{AB}(T_1)(V(T_1)+X_1)$ |
| $Y_A=1,Y_B=0,Y_C=1$ | $p_{101}$ | $\bar{\varphi}_B(T)R(T_0,T_1)$ | $\bar{\varphi}_B(T)R(T_0,T_1)$ |
| $Y_A=0,Y_B=1,Y_C=1$ | $p_{011}$ | $\varphi_B(T)R(T_0,T_1)$ | $\varphi_B(T)R(T_0,T_1)$ |
| $Y_A=1,Y_B=1,Y_C=1$ | $p_{111}$ | $\varphi_{AB}(T)R(T_0,T_1)$ | $\varphi_{AB}(T)R(T_0,T_1)$ |

The risky price is the discounted expectation of the payoffs and is given by

---

*payment rule* is based on current ISDA documentation. The non-defaulting party will pay the full market value of the instrument to the defaulting party if the contract has positive value to the defaulting party.

$$
\begin{aligned}
V(t) = E\Big\{ \Big[ & 1_{(V(T_1) X_1 \geq 0} \Big\langle \big( p_{000}(T_0,T_1) + p_{100}(T_0,T_1)\overline{\varphi}_B(T_1) + p_{010}(T_0,T_1)\varphi_B(T_1) + p_{110}(T_0,T_1)\varphi_{AB}(T_1) \big)(V(T_1) + X_1) \Big| \mathcal{F}_t \Big\rangle \\
& + 1_{(V(T_1)+X_1)<0} \Big\langle \big( p_{000}(T_0,T_1) + p_{100}(T_0,T_1)\varphi_A(T_1) + p_{010}(T_0,T_1)\overline{\varphi}_A(T_1) + p_{110}(T_0,T_1)\varphi_{AB}(T_1) \big)(V(T_1) + X_1) \Big| \mathcal{F}_t \Big\rangle \\
& + \big( p_{001}(T_0,T_1) + p_{101}(T_0,T_1)\overline{\varphi}_B(T_1) + p_{011}(T_0,T_1)\varphi_B(T_1) + p_{111}(T_0,T_1)\varphi_{AB}(T_1) \big) R(T_s,T_1) \Big| \mathcal{F}_t \Big] D(t,T) \Big\} \\
= E\Big\{ \Big[ & O(T_0,T_1)(V(T_1) + X_1) + \Omega(T_0,T_1) R(T_s,T_1) \Big] \Big| \mathcal{F}_t \Big\}
\end{aligned}
\tag{A1a}
$$

where

$$
O(T_0,T_1) = 1_{(V(T_1)+X_1)\geq 0}\,\phi_B(T_0,T_1) + 1_{(V(T_1)+X_1)<0}\,\phi_A(T_0,T_1) \tag{A1b}
$$

$$
\begin{aligned}
\phi_A(T_0,T_1) = \{ & p_A(T_0,T_1) p_B(T_0,T_1) p_C(T_0,T_1) + q_A(T_0,T_1) p_B(T_0,T_1) p_C(T_0,T_1) \varphi_A(T_1) \\
& + p_A(T_0,T_1) q_B(T_0,T_1) p_C(T_0,T_1) \overline{\varphi}_A(T_1) + q_A(T_0,T_1) q_B(T_0,T_1) p_C(T_0,T_1) \varphi_{AB}(T_1) \\
& + p_C(T_0,T_1) \sigma_{AB}(T_0,T_1) \big(1 - \varphi_A(T_1) - \overline{\varphi}_A(T_1) + \varphi_{AB}(T_1)\big) \\
& + \sigma_{AC}(T_0,T_1) \big[ p_B(T_0,T_1)\big(1 - \varphi_A(T_1)\big) + q_B(T_0,T_1)\big(\overline{\varphi}_A(T_1) - \varphi_{AB}(T_1)\big) \big] \\
& + \sigma_{BC}(T_0,T_1) \big[ p_A(T_0,T_1)\big(1 - \overline{\varphi}_A(T_1)\big) + q_A(T_0,T_1)\big(\varphi_A(T_1) - \varphi_{AB}(T_1)\big) \big] \\
& + \theta_{ABC}(T_0,T_1) \big( -1 + \overline{\varphi}_A(T_1) - \varphi_{AB}(T_1) + \varphi_A(T_1) \big) \} D(T_0,T_1)
\end{aligned}
\tag{A1c}
$$

$$
\begin{aligned}
\phi_B(T_0,T_1) = \{ & p_A(T_0,T_1) p_B(T_0,T_1) p_C(T_0,T_1) + q_A(T_0,T_1) p_B(T_0,T_1) p_C(T_0,T_1) \overline{\varphi}_B(T_1) \\
& + p_A(T_1,T_1) q_B(T_0,T_1) p_C(T_0,T_1) \varphi_B(T_1) + q_A(T_0,T_1) q_B(T_0,T_1) p_C(T_0,T_1) \varphi_{AB}(T_1) \\
& + p_C(T_0,T_1) \sigma_{AB}(T_0,T_1) \big(1 - \varphi_B(T_1) - \overline{\varphi}_B(T_1) + \varphi_{AB}(T_1)\big) \\
& + \sigma_{AC}(T_0,T_1) \big[ p_B(T_0,T_1)\big(1 - \overline{\varphi}_B(T_1)\big) + q_B(T_0,T_1)\big(\varphi_B(T_1) - \varphi_{AB}(T_1)\big) \big] \\
& + \sigma_{BC}(T_0,T_1) \big[ p_A(T_0,T_1)\big(1 - \varphi_B(T_1)\big) + q_A(T_0,T_1)\big(\overline{\varphi}_B(T_1) - \varphi_{AB}(T_1)\big) \big] \\
& + \theta_{ABC}(T_0,T_1) \big( -1 + \overline{\varphi}_B(T_1) - \varphi_{AB}(T_1) + \varphi_B(T_1) \big) \} D(T_0,T_1)
\end{aligned}
\tag{A1d}
$$

$$
\begin{aligned}
\Omega(T_0,T) = \{ & p_A(T_0,T) p_B(T_0,T) q_C(T_0,T) + q_A(T_0,T) p_B(T_0,T) q_C(T_0,T) \overline{\varphi}_B(T_1) \\
& + p_A(T_0,T) q_B(T_0,T) q_C(T_0,T) \varphi_B(T_1) + q_A(T_0,T) q_B(T_0,T) q_C(T_0,T) \varphi_{AB}(T_1) \\
& + q_C(T_0,T) \sigma_{AB}(T_0,T) \big(1 - \varphi_B(T_1) - \overline{\varphi}_B(T_1) + \varphi_{AB}(T_1)\big) \\
& - \sigma_{AC}(T_0,T) \big[ p_B(T_0,T)\big(1 - \overline{\varphi}_B(T_1)\big) + q_B(T_0,T)\big(\varphi_B(T_1) - \varphi_{AB}(T_1)\big) \big] \\
& - \sigma_{BC}(T_0,T) \big[ p_A(T_0,T)\big(1 - \varphi_B(T_1)\big) + q_A(T_0,T)\big(\overline{\varphi}_B(T_1) - \varphi_{AB}(T_1)\big) \big] \\
& + \theta_{ABC}(T_0,T) \big( 1 - \overline{\varphi}_B(T_1) + \varphi_{AB}(T_1) - \varphi_B(T_1) \big) \} D(T_0,T)
\end{aligned}
\tag{A1e}
$$

Similarly, we have

$$
V(T_1) = E\Big\{ \big[ O(T_1,T_2)\big(X_2 + V(T_2)\big) + \Omega(T_1,T_2) R(T_1,T_2) \big] \Big| \mathcal{F}_{T_1} \Big\} \tag{A2}
$$

Note that $O(T_0,T_1)$ is $\mathcal{F}_{T_1}$-measurable. By definition, an $\mathcal{F}_{T_1}$-measurable random variable is a random variable whose value is known at time $T_1$. According to *taking out what is known* and *tow*er properties of conditional expectation, we have

$$
\begin{aligned}
V(t) &= E\{[O(T_0,T_1)(X_1+V(T_1))+\Omega(T_0,T_1)R(T_0,T_1)]|\mathcal{F}_t\} = E[O(T_0,T_1)X_1|\mathcal{F}_t] \\
&\quad + E[\Omega(T_0,T_1)R(T_0,T_1)|\mathcal{F}_t] + E\{O(T_0,T_1)E\langle[O(T_1,T_2)(X_2+V(T_2))+\Omega(T_1,T_2)R(T_1,T_2)]|\mathcal{F}_{T_1}\rangle|\mathcal{F}_t\} \\
&= \sum_{i=1}^{2} E\left[\left(\prod_{j=0}^{i-1} O(T_j,T_{j+1})\right)X_i\right] + \sum_{i=1}^{2} E\left[\left(\prod_{j=0}^{i-2} O(T_j,T_{j+1})\right)\Omega(T_{i-1},T_i)R(T_{i-1},T_i)\right] \\
&\quad + E\left[\left(\prod_{j=0}^{1} O(T_j,T_{j+1})\right)V(T_2)|\mathcal{F}_t\right]
\end{aligned}
\tag{A3}
$$

By recursively deriving from $T_2$ forward over $T_m$, where $V(T_m) = X_m$, we have

$$
V(t) = \sum_{i=1}^{m} E\left[\left(\prod_{j=0}^{i-1} O(T_j,T_{j+1})\right)X_i|\mathcal{F}_t\right] + \sum_{i=1}^{m} E\left[\left(\prod_{j=0}^{i-2} O(T_j,T_{j+1})\right)\Omega(T_{i-1},T_i)R(T_{i-1},T_i)|\mathcal{F}_t\right]
\tag{A4}
$$

## References

Arora, Navneet, Priyank Gandhi, and Francis A. Longstaff (2009), "Counterparty credit risk and the credit default swap market," Working paper, UCLA.

Brigo, D., A. Pallavicini, and R. Torresetti (2011), "Credit Models and the Crisis: default cluster dynamics and the Generalized Poisson Loss model," Journal of Credit Risk, 6: 39-81.

Brigo, D., and Capponi, A., 2008, Bilateral counterparty risk valuation with stochastic dynamical models and application to Credit Default Swaps, Working paper.

Collin-Dufresne, P., R. Goldstein, and J. Helwege (2003), "Is credit event risk priced? Modeling contagion via the updating of beliefs," Working paper, Haas School, University of California, Berkeley.

Das, S., D. Duffie, N. Kapadia, and L. Saita (2007), "Common failings: How corporate defaults are correlated," Journal of Finance, 62: 93-117.

Das, S., L. Freed, G. Geng, N. Kapadia (2006), "Correlated default risk," Journal of Fixed Income, 16: 7-32.

Davis, M., and V. Lo (2001), "Infectious defaults," Quantitative Finance, 1: 382-387.

Bielecki, T., I. Cialenco and I. Iyigunler, (2013) "Collateralized CVA valuation with rating triggers and credit migrations," International Journal of Theoretical and Applied Finance, 16 (2).

Crepey, S. (2015) "Bilateral counterparty risk under funding constraints – part II: CVA," Mathematical Finance, 25(1), 23-50.

Deardorff, Alan V. (1982): "The general validity of the Heckscher-Ohlin Theorem," American Economic Review, 72 (4): 683-694.

Duffie, D., A. Eckner, G. Horel, and L. Saita (2009), "Frailty correlated default," Journal of Finance, 64: 2089-2123.

Duffie, D., and K. Singleton (1999), "Modeling term structure of defaultable bonds," Review of Financial Studies, 12: 687-720.

ECB (2009), "Credit default swaps and counterparty risk," European central bank.

Fuijii, M. and A. Takahahsi (2012), "Collateralized CDS and default dependence – Implications for the central clearing," Journal of Credit Risk, 8(3): 97-113.

Giesecke, K. (2004), "Correlated default with incomplete information," Journal of Banking and Finance, 28: 1521-1545.

Gregory, Jon, 2009, Being two-faced over counterparty credit risk, RISK, 22, 86-90.

Hull, J. and A. White (2004), "Valuation of a CDO and a nth to default CDS without Monte Carlo simulation," Journal of Derivatives, 12: 8-23.

Hull, J. and A. White (2014), "Collateral and Credit Issues in Derivatives Pricing," Journal of Credit Risk, 10 (3), 3-28.

ISDA (2013), "ISDA margin survey 2013."

Jarrow, R., and S. Turnbull (1995), "Pricing derivatives on financial securities subject to credit risk," Journal of Finance, 50: 53-85.

Jarrow, R., and F. Yu (2001), "Counterparty risk and the pricing of defaultable securities," Journal of Finance, 56: 1765-99.

Johannes, M. and S. Sundaresan (2007), "The impact of collateralization on swap rates," Journal of Finance, 62: 383-410.

J. P. Morgan (1999), "The J. P. Morgan guide to credit derivatives," Risk Publications.

Koopman, S., A. Lucas, and B. Schwaab (2011), "Modeling frailty-correlated defaults using many macroeconomic covariates," Journal of Econometrics, 162: 312-325.

Lando, D. and M. Nielsen (2010), "Correlation in corporate defaults: contagion or conditional independence?" Journal of Financial Intermediation, 19: 355-372.

Laurent, J. and J. Gregory (2005), “Basket default swaps, CDOs and factor copulas,” Journal of Risk, 7: 03-122.

Li, D. (2000), “On default correlation: A copula function approach,” Journal of Fixed Income, 9: 43-54.

Lipton, A., and Sepp, A., 2009, Credit value adjustment for credit default swaps via the structural default model, Journal of Credit Risk, 5(2), 123-146.

Longstaff, F., and E. Schwartz (2001): “Valuing American options by simulation: a simple least-squares approach,” The Review of Financial Studies, 14 (1): 113-147.

Madan, D., and H. Unal (1998), “Pricing the risks of default,” Review of Derivatives Research, 2: 121-160.

Piterbarg, V. (2010), “Funding beyond discounting: collateral agreements and derivatives pricing,” Risk Magazine, 2010 (2): 97-102.

Pykhtin, Michael, and Steven Zhu, 2007, A guide to modeling counterparty credit risk, GARP Risk Review, July/August, 16-22.

Teugels, J. (1990), “Some representations of the multivariate Bernoulli and binomial distributions,” Journal of Multivariate Analysis, 32: 256-268.

Zhang, G., and P. Jorion, (2007), “Good and bad credit contagion: Evidence from credit default swaps,” *Journal of Financial Economics* 84: 860–883.